\documentclass[%
 reprint,
 superscriptaddress,
 amsmath,amssymb,
 aps,
 prb,
]{revtex4-1}

\usepackage{graphicx}% Include figure files
\usepackage{dcolumn}% Align table columns on decimal point
\usepackage{bm}% bold math
\usepackage{mathtools}
\usepackage{color}
\usepackage{cases}
\usepackage{booktabs} % horizontal rules

\newcommand{\dd}[2]{\ensuremath{\frac{\mathrm{d}#1}{\mathrm{d}#2}}}

\begin{document}

\preprint{APS/123-QED}

\title{Response to noise of a spin transfer vortex based nano-oscillator}% Force line breaks with \\

\author{Eva Grimaldi}
\affiliation{%
 Unit\'e Mixte de Physique CNRS/Thales and Universit\'e Paris-Sud 11, 1 Ave. A. Fresnel, 91767 Palaiseau, France
}
\affiliation{%
 CNES, 1 Ave. Edouard Belin, 31400 Toulouse, France
}
\author{Antoine Dussaux}\altaffiliation{Department of Physics, ETH Zurich, Schafmattstrasse 16, 8093 Zurich, Switzerland}
\affiliation{%
 Unit\'e Mixte de Physique CNRS/Thales and Universit\'e Paris-Sud 11, 1 Ave. A. Fresnel, 91767 Palaiseau, France
}
\author{Paolo Bortolotti}%
\affiliation{%
 Unit\'e Mixte de Physique CNRS/Thales and Universit\'e Paris-Sud 11, 1 Ave. A. Fresnel, 91767 Palaiseau, France
}
\affiliation{Thales Research and Technology, 1 Ave. A. Fresnel, 91767 Palaiseau, France}
\author{Julie Grollier}\affiliation{%
 Unit\'e Mixte de Physique CNRS/Thales and Universit\'e Paris-Sud 11, 1 Ave. A. Fresnel, 91767 Palaiseau, France
}%}%
\author{Gr\'egoire Pillet}
\affiliation{Thales Research and Technology, 1 Ave. A. Fresnel, 91767 Palaiseau, France}
\author{Akio Fukushima}
\author{Hitoshi Kubota}
\author{Kay Yakushiji}
\author{Shinji Yuasa}
\affiliation{
 National Institute of Advanced Industrial Science and Technology (AIST) 1-1-1 Umezono, Tsukuba, Ibaraki 305-8568, Japan
}%
\author{Vincent Cros}%
\email{vincent.cros@thalesgroup.com}
\affiliation{%
 Unit\'e Mixte de Physique CNRS/Thales and Universit\'e Paris-Sud 11, 1 Ave. A. Fresnel, 91767 Palaiseau, France
}%

\date{\today}% It is always \today, today,

\begin{abstract}
We investigate experimentally and analytically the impact of thermal noise on the sustained gyrotropic mode of vortex magnetization in spin transfer nano-oscillators and its consequence on the linewidth broadening due to the different nonlinear contributions. Performing some time domain measurements, we are able to extract separately the phase noise and the amplitude noise at room temperature for several values of dc current and perpendicular field. For a theoretical description of the experiments, we extend the general model of nonlinear auto-oscillators to the case of vortex core dynamics and provide some analytical expressions of the response-to-noise of the system as the coupling coefficient between the phase and the amplitude of the vortex core dynamics due to the nonlinearities. From the analysis of our experimental results, we demonstrate the major role of the amplitude-to-phase noise conversion on the linewidth broadening, and propose some solutions to improve the spectral coherence of vortex based spin transfer nano-oscillators.

\end{abstract}

\pacs{Valid PACS appear here}% PACS, the Physics and Astronomy
\maketitle

%\tableofcontents

\section*{Introduction}

Since the discovery of spin transfer induced dynamics, experimental and analytical studies have led to a much better understanding of the microscopic mechanism of the transfer of angular momentum in nanostructures and the magnetization dynamics that are generated\citep{berger_emission_1996,slonczewskI_current-driven_1996}. One of the objectives is to improve the characteristic of the rf emitted power associated to the magnetization dynamics, to lead to the development of a new type of integrated and tunable frequency source, the so called spin transfer nano-oscillator (STNO). Such oscillators are based on the conversion of magnetization dynamics into voltage oscillations. Being nanoscale and with an rf frequency from few tens of MHz to few tens of GHz makes them good candidates for telecommunication applications\citep{tulapurkar_spin-torque_2005,katine_device_2008}. In the scope of understanding both the magnetization dynamics and improvements of rf features, several issues have been addressed: how to excite magnetic modes with a spin polarized current\citep{kiselev_microwave_2003,rippard_direct-current_2004,mistral_current-driven_2008}, how to improve the STNO output power and decrease critical current densities\citep{dussaux_large_2010} and, eventually, how the spectrum purity is affected by thermal fluctuations and nonlinearities\citep{mistral_current-driven_2006,kudo_amplitude-phase_2009,georges_origin_2009}.

The fundamental mode of vortex magnetization\citep{cowburn_single-domain_1999,shinjo_magnetic_2000} called the gyrotropic mode has been intensively studied in STNOs. The magnetization oscillations can be reduced to the dynamics of the vortex core\citep{thiele_steady-state_1973} which give rise to large amplitude magnetization oscillations with frequencies from 100~MHz up to 2~GHz. In general, it exhibits a relatively small linewidth compared to STNOs based on other excited mode e.g. a uniform precession. However, the understanding of the linewidth broadening is still a challenge\citep{sankey_mechanisms_2005,bortolotti_temperature_2012}. The vortex magnetization spin transfer torque induced dynamics in STNOs can be described by its amplitude and phase. As the magnetization dynamics are affected by the exchange of energy with the thermal bath, amplitude and phase get blurred resulting in amplitude and phase noise, and a finite linewidth is measured. In order to understand the origin of this linewidth broadening, in Section~\ref{sec:section1} we present the experimental results corresponding to the amplitude and phase noise of the gyrotropic motion at room temperature from single shot time domain measurements\citep{picinbono_instantaneous_1997,bianchini_direct_2010,quinsat_amplitude_2010}. In Section~\ref{sec:section2}, we develop a general auto-oscillator model\citep{mizushima_analytical_2007,tiberkevich_microwave_2007,kim_generation_2008,slavin_nonlinear_2009,silva_theory_2010} for our case of interest that corresponds to the case of vortex magnetization dynamics. One of the important results is that we provide some analytical expression of the main parameters that describe the response-to-noise of the vortex based oscillator. Finally, in Section~\ref{sec:section3} we compare experiments and theoretical predictions allowing us to validate the model in the case of vortex dynamics. From this comparison, we can determine the main source of noise. Consequently we propose some solutions to improve furthermore the spectral coherence of vortex based STNOs in order to reach the characteristics required for integrated rf nanosources.

\section{Experiments : frequency vs time domain}
\label{sec:section1}
\begin{figure}[h]
\includegraphics[width=8.4cm]{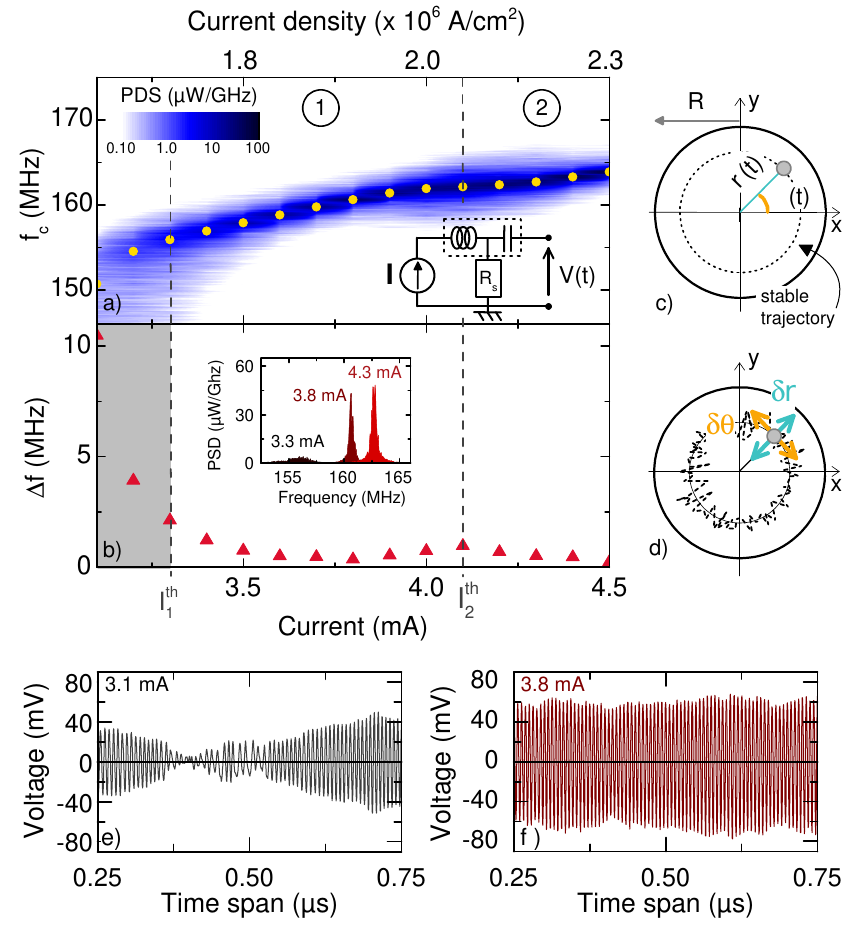}%
\caption{\label{fig:figure00} Rf characteristics of a vortex based STNO measured for $H_{\bot}$\;=\;4.4~kOe: (a) Color scale of the PSD vs current and (yellow dots) carrier frequency $f_c$ extracted from fit. Inset: Schematic of the measurement set-up. The $R_s$ resistance represents the STNO resistance, and $V(t)$ is the input measurement voltage. (b) $\Delta\!f$ vs current. Inset : Voltage PSD measured for 3.3, 3.8 and 4.3~mA. (c) Schematic of the vortex core sustained trajectory described by amplitude $r(t)$ and phase $\theta(t)$. (d) Schematic of vortex core dynamics affected by amplitude noise $\delta r$ and phase noise $\delta\theta$. (e) An example of time trace below $I_{1}^{th}$ at 3.1~mA. (f) An example of time trace above $I_{1}^{th}$ at 3.8~mA. Time traces are measured with a 30~dB amplifier.}
\end{figure}

The studied samples are circular tunnel junctions made of a layered stack //\,Synthetic antiferromagnet\,(SAF)\,/\,MgO\,(1.075)\,/\,NiFe\,(5) (with thickness in nm) of radius R\,=\,250~nm. The NiFe free layer has a vortex distribution magnetization with the vortex core polarity set by the external out-of-plane field $H_{\bot}$ direction and the chirality set by the dc current induced Oersted field. The SAF is composed of PtMn\,(15)\,/\,CoFe\,(2.5)\,/\,Ru\,(0.85)\,/\,CoFeB\,(3) and its top layer magnetization is uniform and lies in the film plane. The vortex core position is converted into a voltage via the tunnel magnetoresistance (TMR\,=\,15\% at room temperature). When an external magnetic field $H_{\bot}$ is applied perpendicular to the film plane, the SAF top layer magnetization is tilted out of the plane. This gives the necessary spin polarization component of the spin transfer torque to sustain the gyrotropic motion with an uniformly spin polarized current\citep{khvalkovskiy_vortex_2009,dussaux_large_2010}. We have performed a comprehensive study at different field values. In this paper, we focus on the data that have been measured at $H_{\bot}$\;=\;4.4~kOe, but similar conclusions have been obtained for the other field values. When a dc current $I$ is injected ($I$\,$>$\,$0$ is defined for electrons flowing from the free layer to the SAF layer), the vortex core oscillates around the dot center along a circular trajectory. Using a spectrum analyzer, we measure the STNO voltage power spectral density plotted in color scale in Fig.~\ref{fig:figure00}(a). The measurement set-up is shown in the inset in Fig.~\ref{fig:figure00}(a). Finally we note that all the measurements are performed at room temperature.

The vortex core dynamics are described by the evolution in time of the orbit radius r(t) and the phase $\theta(t)$ as schematized in Fig.~\ref{fig:figure00}(c). Hence, we define the normalized orbit radius $s(t)$\,=\,$r(t)/R$) and the oscillation frequency $\dot{\theta}(t)/2\pi$ . Experimentally, we extract the voltage rf properties that are the carrier frequency $f_c$, the integrated power and the full-width at half-maximum $\Delta\!f$ from a lorentzian fit of the power spectral density of the measured voltage. These quantities are linked to the vortex core dynamics as reported in Appendix~\ref{Ap:app1}: $f_c$ corresponds to the oscillation frequency of the core $\dot{\theta}$/2$\pi$, the integrated power is proportional to the normalized orbit radius of the core $s_0$ and the linewidth $\Delta\!f$ is related to the orbit stability. The stable orbit is perturbed by noise through amplitude noise $\delta$s(t)\,=\,$\delta$r(t)/R (radial perturbations) and phase noise $\delta\theta$ (orthoradial perturbations along the sustained trajectory) as sketched in Fig.~\ref{fig:figure00}(d). The conversion of the vortex dynamic noise into the voltage noise is detailed in Appendix~\ref{Ap:app1} and it is assumed that the voltage normalized amplitude noise $\delta\alpha$ (resp. the voltage phase noise $\delta\phi$) is equal to the magnetization normalized amplitude noise $\delta s(t)/s_0$ (resp. the magnetization phase noise $\delta\theta$).

In Fig.~\ref{fig:figure00}(a-b), we plot the evolution of the frequency carrier $f_c$ and of the linewidth $\Delta\!f$ of the rf signal as a function of the injected current $I$. Below the current $I_{1}^{th}$\,=\,3.3~mA, we measure a large $\Delta\!f$ corresponding to large amplitude vortex oscillations but these are in fact not continuously sustained over long timescales as plotted in Fig.~\ref{fig:figure00}(e). Above $I_{1}^{th}$ up to $I_{2}^{th}$\,=\,4.1~mA, defined as the region~\textcircled{\raisebox{-1.0pt}{\hspace*{0.0pt}1}}, the vortex core exhibits large amplitude sustained oscillations induced by spin transfer (see a typical time trace in this region in Fig.~\ref{fig:figure00}(f)). In this region, increasing the current corresponds to a decrease in $\Delta\!f$ from 2~MHz down to 360~kHz at $I$\,=\,3.8~mA. For current values close to $I_{2}^{th}$, we observe an increase of $\Delta\!f$ and a flattening of the frequency carrier evolution: the region~\textcircled{\raisebox{-1.0pt}{\hspace*{0.0pt}2}} defined for high currents, above $I_{2}^{th}$, corresponds to large amplitude oscillations ($50\%$ of the radius) and large velocity ($\simeq$ 130 m/s) where a substantial change of the vortex core shape occurs. For this region, our analytical description is no longer valid. The vicinity of $I_{2}^{th}$ corresponds most probably to a transition region towards another dynamical regime, that implies a change in the linewidth evolution\citep{muduli_decoherence_2012}. In the following, we focus on the regime of vortex oscillations corresponding to the region~\textcircled{\raisebox{-1.0pt}{\hspace*{0.0pt}1}} where a single mode exists. Under the assumption of a single mode excited by spin transfer torque, both the integrated power and carrier frequency evolution have been previously well described analytically\citep{dussaux_field_2012}. However, there are still some open questions because recent works\citep{schneider_temperature_2009,bortolotti_temperature_2012,sierra_influence_2012} showed that the linewidth $\Delta\!f$ decreases linearly with decreasing the temperature down to about 100~K as expected from fluctuation-dissipation theory but then saturates. Thus, the detailed mechanisms at the origin of the linewidth broadening in such vortex based oscillators, as well as the evolution of the linewidth $\Delta\!f$, remain to be understood.

In the following, we investigate the linewidth $\Delta\!f$ broadening under conditions for the applied perpendicular field and dc current for which the sustained oscillations are obtained (region~\textcircled{\raisebox{-1.0pt}{\hspace*{0.0pt}1}} in Fig.~\ref{fig:figure00}(a-b)). The spectral quality of the vortex based STNO can be investigated quantitatively by separating the voltage amplitude noise $\delta\alpha(t)$ and the voltage phase noise $\delta\phi(t)$ from the voltage output signal $V(t)$\,=\,$V_0 (1+\delta\alpha(t)) \cos(2\pi ft+\delta\phi(t))$\citep{keller_nonwhite_2010}. The phase noise has been previously measured for vortex based STNOs by Keller \textit{et al.}\citep{keller_time_2009}. However, because of the nonlinear nature of STNOs, the amplitude-phase coupling makes the additional measurement of amplitude noise of significant importance to fully analyze the spectral coherence of vortex based oscillators. A signal analysis called the Hilbert Transform Method was applied to characterize the experimental output signal of STNOs based on uniform magnetization dynamics\citep{picinbono_instantaneous_1997,bianchini_direct_2010,quinsat_amplitude_2010}.

\begin{figure}[h]
\includegraphics[width=8.3cm]{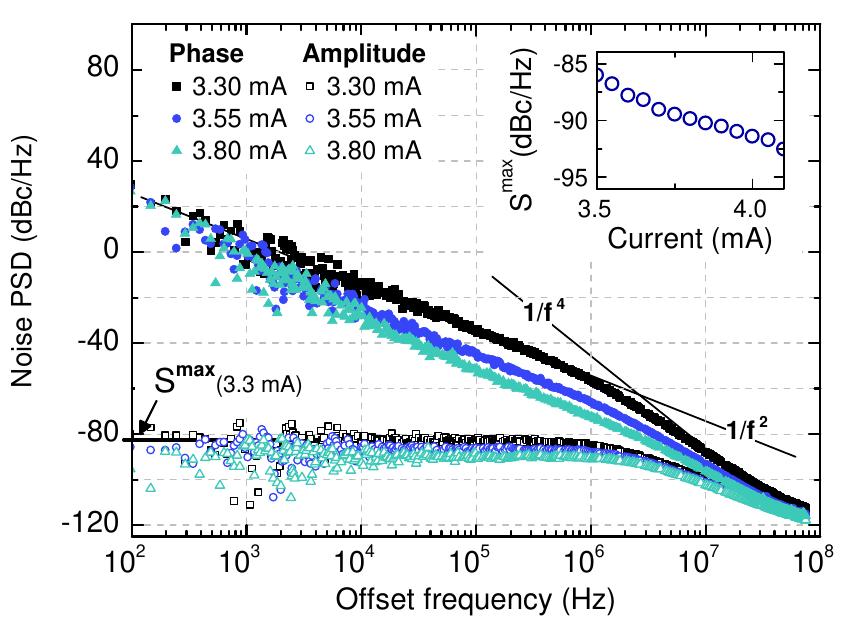}%
\caption{\label{fig:figure0} Amplitude (empty dots) and phase (full dots) noise PSD at $H_{\bot}$\,=\,4.4~kOe obtained for different values of current $I$: 3.30~mA (black squares), 3.55~mA (blue circles) and 3.80~mA (green triangles) that correspond respectively to the STNO output voltage integrated power of 11~nW, 15~nW and 23~nW. Inset: Amplitude noise maximum level versus current $I$.}
\end{figure}

To perform the analysis of phase and amplitude noise, we measured a 20.5~ms long STNO output voltage time trace with a single shot oscilloscope over $41\times10^6$ points at a sampling rate of 2~GSa/s. The STNO output voltage V(t) is amplified before the oscilloscope input (gain of 30~dB and noise figure of 1.8 dB). An internal filter of the oscilloscope is applied to remove frequencies higher than the fundamental gyrotropic mode. By applying the Hilbert Transform Method, we extract the phase and amplitude noise. In Fig.~\ref{fig:figure0}, we plot in logarithmic scale the power spectral densitiy (PSD) of amplitude and phase noise as a function of the offset frequency from the carrier frequency $(f-f_c)$ for 3.3, 3.55 and 3.8~mA. The noise PSDs are limited by the electrical Johnson-Nyquist noise floor increased by the amplifier noise figure around $-120$~dBc/Hz at offset frequencies above $10^7$-$10^8$~Hz. The amplitude noise is much lower than the phase noise below 1~MHz demonstrating that the phase noise is indeed the main contribution to the linewidth $\Delta\!f$ in vortex based STNOs, as it is the case for most oscillating systems in nature. For all currents, the amplitude noise PSD shows a lorentzian distribution. At small current $I$, the phase noise PSD presents a $1/f^2$ dependence at small offset frequencies and a $1/f^n$ (with $2<n\leq4$) at higher offset frequencies. When the current increases, the amplitude noise maximum level $S^{max}_{\delta\alpha}$ decreases (see the inset in Fig.~\ref{fig:figure0}): the STNO output voltage $V_0$ increases and the relative amplitude noise, $\delta V/V_0\,=\,\delta\alpha$, decreases. Thus at higher currents, the amplitude noise decreases, and the phase noise PSD tends to a single $1/f^2$ distribution over the whole frequency range which is equivalent to a white frequency noise. This trend of phase noise PSD is a signature of the amplitude-phase coupling due to the nonlinearities and are discussed in the following section.
\section{Vortex magnetization dynamics in the framework of the nonlinear auto-oscillator}
\label{sec:section2}
The amplitude and phase noise was analyzed with a general model proposed for nonlinear auto-oscillators\citep{mizushima_analytical_2007,tiberkevich_microwave_2007,kim_generation_2008,slavin_nonlinear_2009,silva_theory_2010} that takes into account the nonlinear behavior of the system and showed that the linewidth is due to both thermally driven phase noise and amplitude-phase coupling. In this paper, we demonstrate that such approach is relevant for vortex based STNOs too.

To understand the linewidth broadening of the STNO and the phase noise distribution, the magnetization oscillations must be considered and studied as non deterministic dynamics. A general model for nonlinear auto-oscillators\citep{tiberkevich_microwave_2007,slavin_nonlinear_2009} developed by A. Slavin and V. Tiberkevich considered that the deterministic dynamics of a nonlinear oscillator can be described by the equation of the complex amplitude $c(t)$:
\begin{equation}
	\label{eq:equ2} \dd{c}{t}\, + \, i2\pi f(p) c \, + \, \Gamma_+ (p) c \, - \, \Gamma_- (p) c \, = \, 0
\end{equation}
with $f$ the instantaneous frequency of the oscillator, $\Gamma_+$ the positive damping rate that represents the losses of the system and $\Gamma_-$ the negative damping rate that represents the gain of the system. These three parameters depend on the oscillation power $p\,=\,|c|^2$. Solving the equation $\Gamma_-(p)$\,=\,$\Gamma_+(p)$  gives the stationary solution for the oscillation power $p_0$. Moreover, this general model is a powerful tool in the case of non-deterministic dynamics induced by temperature\citep{kim_stochastic_2006,kim_generation_2008,kim_line_2008,tiberkevich_temperature_2008}: the response-to-noise of the system around its stable trajectory can be described with three key-parameters, namely the damping rate for small power deviations from stationary solution $\pi f_p\,=\,(\dd{\Gamma_+}{p}(p_0) -\dd{\Gamma_-}{p}(p_0))p_0$ that gives the rate at which the system goes back to the stable trajectory when perturbed, the auto-oscillator generation linewidth in the linear regime $2\Delta\!f_0\,=\,\frac{1}{2\pi}\frac{k_BT}{\epsilon(p_0)}\Gamma_+(p_0)$ (with $\epsilon(p)$ the energy of the system) which is the intrinsic phase noise due to the thermal bath and the normalized dimensionless nonlinear frequency shift $\nu\,=\,\frac{N}{\pi f_p}p_0$ (with the nonlinear frequency shift coefficient $N\,=\,2\pi\dd{f(p)}{p}$) that quantifies coupling between phase and amplitude. Recently, it has been demonstrated that this general model could be adapted to STNOs with uniform magnetization dynamics: the predicted results for a white noise affecting the magnetization dynamics described by Eq.~(\ref{eq:equ2}) were in good agreement with the phase and amplitude noise obtained experimentally\citep{quinsat_amplitude_2010,sierra_influence_2012,quinsat_linewidth_2012}.

We have shown recently\citep{dussaux_field_2012} that for vortex based STNOs, the STT induced gyrotropic motion of the vortex core can be very well described by the Thiele equation\citep{thiele_steady-state_1973} considering the second orders terms of damping and confinement. This allows a deterministic description of the vortex magnetization dynamic frequency and amplitude through the polar coordinates of the vortex core $(s(t),\theta(t))$ in the normalized disk  (see Fig.~\ref{fig:figure00}(c)):
\begin{equation}
\label{eq:equ3} 
\begin{dcases}
	\dd{\theta}{t} \, = \, \frac{\kappa}{G}(1 + \zeta s^2)\\
	\dd{s}{t} \, = \,  \frac{D\kappa}{G^2}s\left(\frac{a_jIG}{D\kappa\pi R^2}-1+(\zeta+\xi)s^2\right)\\
\end{dcases}
\end{equation}
with $R$ the radius of the ferromagnetic disk, $G$ the gyrovector magnitude, $D(1+\xi s^2)$ the damping coefficient with $D$ its linear part and $\xi$ its nonlinearity factor, $a_j$ the spin transfer torque efficiency, $I$ the current and $\kappa(1+\zeta s^2)$ the confinement stiffness with $\kappa$ its linear part and $\zeta$ its nonlinearity factor. The confinement stiffness is expressed in terms of the magnetostatic confinement $\kappa_{ms}$, the confinement due to the Zeeman interaction of the in-plane vortex magnetization with the Oersted field $\kappa_{oe}I/\pi R^2$ and the nonlinearity factors $\kappa'_{ms}$ and $\kappa'_{oe}I/\pi R^2$ where $\kappa=\kappa_{ms}+\kappa_{oe}I/\pi R^2$ and  $\zeta\,=\,\frac{(\kappa'_{ms}+\kappa'_{oe}I/\pi R^2)}{(\kappa_{ms}+\kappa_{oe}I/\pi R^2)}$ (see Appendix~\ref{Ap:app3} for more details). This description was shown to be in good agreement with the evolution with current and field observed in micromagnetic simulations\citep{dussaux_field_2012} and with experimental results\citep{a._dussaux_unpublished_2013}.

Here we derive the general auto-oscillator equation (\ref{eq:equ2}) for the particular case of vortex magnetization dynamics in STNOs in the presence of a white noise. The auto-oscillator equation~(\ref{eq:equ2}) and the equations of phase and amplitude dynamics~(\ref{eq:equ3}) are equivalent when writing $c (t)\,=\,s(t) e^{-i \theta(t)}$, where the vortex oscillation power is $p(t)\,=\,s(t)^2$. By identification, we deduce the damping rates specific to the vortex dynamics in a STNO:  
\begin{equation}
\label{eq:equ5}
\begin{dcases}
	\Gamma_+ (p) \,= \, \frac{D\kappa}{G^2} \left(1+(\zeta+\xi)p \right) \\
	\Gamma_- (p) \,= \, \frac{a_jI}{G\pi R^2}
\end{dcases}
\end{equation}

The negative damping rate $\Gamma_-$ only depends on the injected current $I$ and is independent on the power $p$. Thus the existence of a stable solution $p_0$ that verifies $\Gamma_-(p_0)\,=\,\Gamma_+(p_0)$ for different currents is allowed due to adjustment of the confinement (proportional to $I$) and of the nonlinearities (proportional to $p_0$). The resulting stable solution is a perfect circular trajectory of the vortex core:
\begin{equation}
\label{eq:equp0}
	p_0 \,= \, {s_0}^2\,= \,\frac{\frac{a_jIG}{D\kappa\pi R^2}-1}{\zeta+\xi}\\
\end{equation} 
and has an energy $\epsilon(p_0)\,=\,\frac{1}{2}\kappa R^2p_0$. 

An important feature specific to vortex based STNOs has to be emphasized and compared to uniform magnetization based STNOs. For uniform based STNOs, the relatively large tunability comes from nonlinearities through the nonlinear frequency shift coefficient $N$\citep{tiberkevich_microwave_2007}. As a consequence, the linewidth is strongly broadened due these nonlinearities\citep{georges_origin_2009}. In contrast, for vortex based STNOs, there exists two origins of the frequency tunability which arises from both the confinement term\citep{choi_understanding_2008,dussaux_field_2012,a._hamadeh_origin_2013} $\frac{\kappa_{oe}I}{\pi R^2}$ and the nonlinear frequency shift coefficient defined as:
\begin{equation}
N\,=\,\zeta\frac{\kappa}{G}\\
\end{equation}
due to the nonlinearity of the confinement. Thus vortex based STNOs present a relatively large tunability (10~MHz/mA) with a small nonlinear frequency shift coefficient, $N$, resulting in a narrow linewidth $\Delta\!f$.

To describe the effects of thermal fluctuations on the vortex core dynamics, we express analytically the parameters that govern the response to noise of the nonlinear vortex based oscillator:
\begin{numcases}{} 
	\pi f_p \,= \,\frac{a_j I}{G\pi R^2}-\frac{D\kappa(I)}{G^2}\label{eq:equ40} \\
	2\Delta\!f_0 \, = \, \frac{k_BT}{\pi R^2 p_0}\frac{D}{G^2}\left( 1+(\zeta+\xi)p_0 \right) \label{eq:equ41} \\
	\nu \, = \, \frac{\zeta}{\frac{D}{G} (\zeta+\xi)} \label{eq:equ42}
\end{numcases}

Finally, similarly to Quinsat \textit{et al.}\citep{quinsat_amplitude_2010} for the case of uniformly magnetized STNOs, we can express the PSD of phase noise $\delta \theta$ and of normalized amplitude noise $\delta s/s_0$ of the vortex magnetization dynamics: 
\begin{numcases}{}
	S_{\delta s/s_0} \,= \, \frac{\Delta\!f_0}{2 \pi} \frac{1}{f^2 + f_p^2} \label{eq:equ6}\\
	S_{\delta \theta} \,= \, \frac{\Delta\!f_0}{\pi f^2}+ 2\nu^2\frac{{f_p}^2}{f^2} S_{\delta s/s_0} \label{eq:equ7}
\end{numcases}
In summary, the phase noise of the vortex magnetization dynamics (\ref{eq:equ7}) has two contributions: the intrinsic phase noise that originates directly from a pure thermal phase noise proportional to $\Delta\!f_0$ and the contribution that comes from the conversion of the amplitude noise into phase noise through the coupling parameter $\nu f_p$.

\section{Results and discussions}
\label{sec:section3}
\begin{figure}[h]
\includegraphics[width=9.5cm]{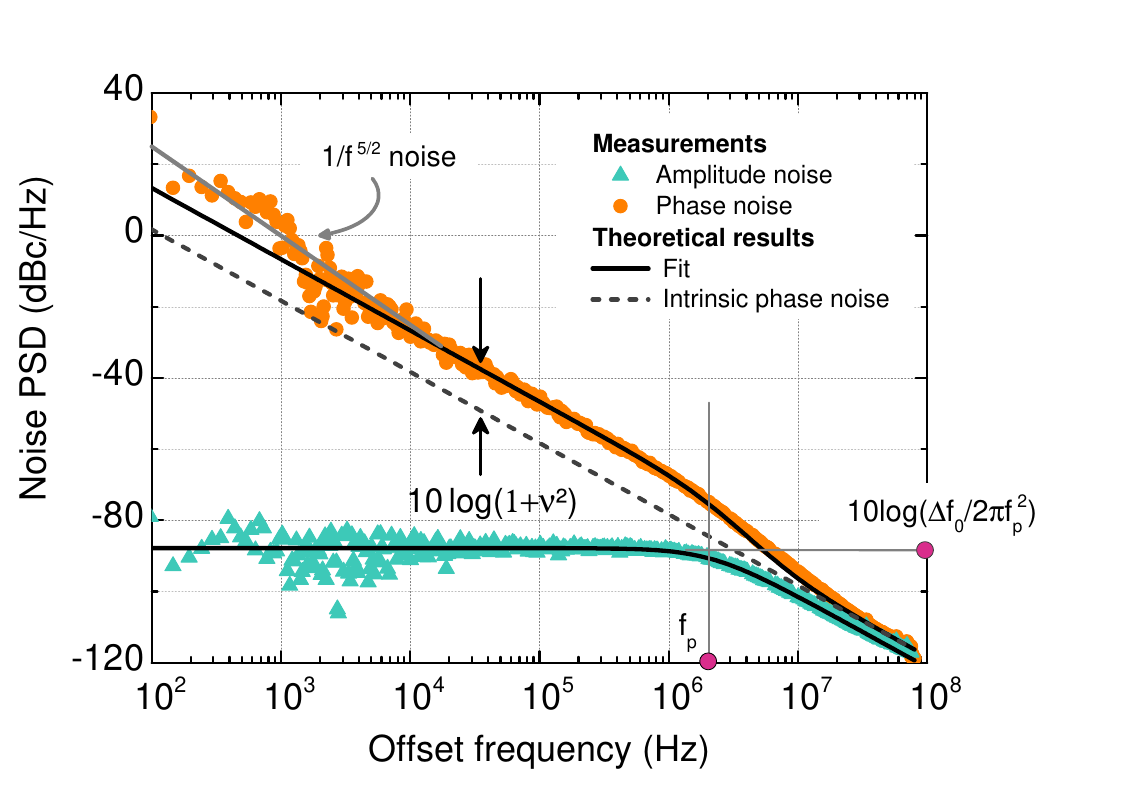}%
\caption{\label{fig:figure1} Phase (orange dots) and amplitude (green triangles) noise PSD in dBc/Hz for $I$\,=\,3.6~mA and $H_{\bot}$\,=\,4.4~kOe. Solid lines correspond to the theoretical amplitude and phase noise PSD from Eq.~(\ref{eq:equ6}) and (\ref{eq:equ7}) for the response-to-noise parameters $f_{p}\,=\,2.1$~MHz, $\Delta\!f_{0}\,=\,46$~kHz and $\nu\,=\,3.7$. The dotted line corresponds to the intrinsic phase noise $\Delta\!f_0 /\pi f^2$. The amplitude-to-phase noise conversion corresponds to an increase of the phase noise of 12~dB.}
\end{figure}
In this section we analyze the noise PSD in the light of the framework described in the previous section. We present the analysis of the experimental data obtained for several currents. In Fig.~\ref{fig:figure1}, the PSD of amplitude noise and phase noise measured for $I$\,=\,3.6~mA are plotted (see symbols). The black lines are fits to the Eq.~(\ref{eq:equ6}) and (\ref{eq:equ7}): the amplitude noise is fitted with Eq.~(\ref{eq:equ6}) to extract the parameters $f_p$ and $\Delta\!f_0$. Then, by injecting these parameters into Eq.~(\ref{eq:equ7}), the phase noise is fitted giving the normalized dimensionless nonlinear frequency shift $\nu$. Note that at a small offset frequency, the phase noise experimental data deviate from the analytical prediction (see gray line of $1/f^{5/2}$ slope). Despite the lack of accuracy at small offset frequencies due to the measurement method, this additional low frequency noise might comes from the MgO barrier\citep{arakawa_low-frequency_2012} but is beyond the scope of our analysis as only noise coming from the magnetization dynamics is considered. Moreover, in region~\textcircled{\raisebox{-1.0pt}{\hspace*{0.0pt}2}}, the amplitude and phase noise PSD can be fitted with the equations Eq.~(\ref{eq:equ7}) and Eq.~(\ref{eq:equ6}) but for currents higher than 5.0~mA, an additional $1/{f}^2$ noise is observed on the phase noise. This additional noise can not be explained by the proposed model and is still under investigation. Measurements performed at other field values give qualitatively identical results.

\begin{figure}[h]
\includegraphics{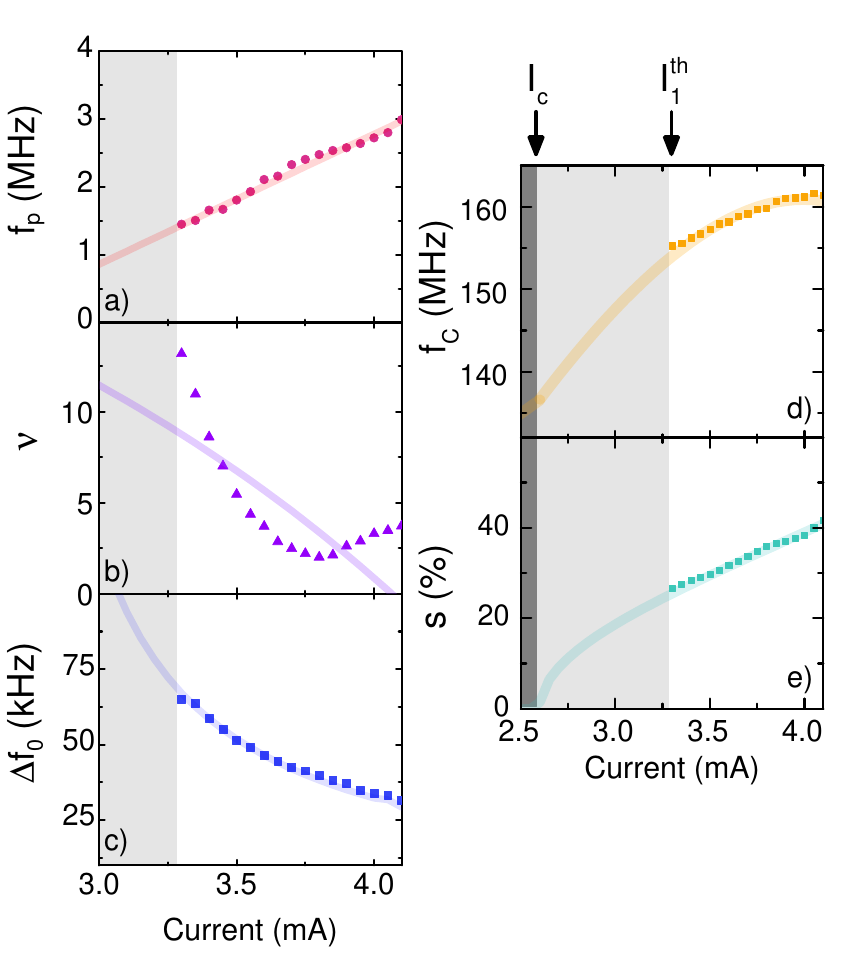}%
\caption{\label{fig:figure2} Response-to-noise parameters versus current for $H_{\bot}$\,=\,4.4~kOe: f$_{p}$ (a), $\nu$ (b) and $\Delta$f$_{0}$ (c). Vortex dynamic features versus current: carrier frequency (d) and normalized orbit radius (e). The symbols correspond to measurement data, and the solid lines to theoretical curves.}
\end{figure}

In Fig.~\ref{fig:figure2}, we plot the response-to-noise parameters $f_{p}$, $\Delta\!f_{0}$ and $\nu$ (symbols) extracted for different values of current in region~\textcircled{\raisebox{-1.0pt}{\hspace*{0.0pt}1}} with the method presented above. 
The damping rate for small power deviations $\pi f_{p}$ is of the order of 1~MHz and increases linearly with current. The normalized dimensionless nonlinear frequency shift $\nu$ decreases sharply down to 2 at 3.8~mA and then increases as getting close to the region~\textcircled{\raisebox{-1.0pt}{\hspace*{0.0pt}2}}. The generation linewidth in the linear regime $2\Delta\!f_{0}$ is of order 10~kHz and decreases with current down to 64~kHz. The evolution of $f_{p}$ and $\Delta\!f_{0}$ with current leads to a decrease of $\frac{\Delta\!f_0}{2\pi {f_p}^2}$, which corresponds to the amplitude noise PSD at small offset frequency referred as $S^{max}_{\delta\alpha}$ in the inset Fig.~\ref{fig:figure0}.

We compare the evolution of the measured response-to-noise parameters to their theoretical evolutions with current $I$ given by Eq.~(\ref{eq:equ40}), (\ref{eq:equ41}) and (\ref{eq:equ42}). 
\begin{table}[h]
\caption{\label{tab:tabParaMagn} Vortex dynamics parameters with $R\,=\,250~$nm, $T\,=\,300~$K and $H_{\bot}$\,=\,4.4~kOe: extracted from the fit of the response-to-noise parameters from Eq.~(\ref{eq:equ40}), (\ref{eq:equ41}) and (\ref{eq:equ42}) and theoretical values assuming the analytical expressions in Appendix~\ref{Ap:app3}.}
\begin{ruledtabular}
\centering
\begin{tabular}{ccc} \toprule
		{} & From fits & From analytical expressions\\
		\hline
    {$\kappa_{ms}$ (kg.s$^{-2}$)} & $4.4\times10^{-5}$ & $4.1\times10^{-5}$\\
   	{$\kappa_{oe}$ (kg.m$^2$.A$^{-1}$.s$^{-2}$)} & $15\times10^{-16}$ & $7.2\times10^{-16}$\\
		{G (kg.rad$^{-1}$.s$^{-1}$)} & $7.4\times10^{-14}$ & $5.3\times10^{-14}$\\
		{D (kg.rad$^{-1}$.s$^{-1}$)} & $19\times10^{-16}$ & $8.8\times10^{-16}$\\
		{a$_j$ (kg.m$^2$.A$^{-1}$.s$^{-2}$)} & $13\times10^{-17}$ & $6.6\times10^{-17}$\\
		{$\kappa'_{ms}$ (kg.s$^{-2}$)} & $2.2\times10^{-4}$ & $0.1\times10^{-4}$\\
		{$\kappa'_{oe}$ (kg.m$^2$.A$^{-1}$.s$^{-2}$)} & $-1.1\times10^{-14}$ & $-3.6\times10^{-16}$\\
		{$\xi$} & $2.0$ & $0.6$\\
\end{tabular}
\end{ruledtabular}
\end{table}
The experimental data (carrier frequency $f_c$, normalized amplitude $s_0$ and response-to-noise parameters $f_{p}$, $\Delta\!f_{0}$ and $\nu$) are fitted with the Eq.~(\ref{eq:equ3}), (\ref{eq:equp0}), (\ref{eq:equ40}), (\ref{eq:equ41}) and (\ref{eq:equ42}) as detailed in Appendix~\ref{Ap:app2} (see solid lines in Fig.~\ref{fig:figure2}) with good agreement except for the normalized dimensionless nonlinear frequency shift $\nu$. From the fitting parameters, we derive the quantities describing the vortex gyrotropic dynamics and list them in Table.~\ref{tab:tabParaMagn}. They reproduce well the gyrotropic dynamic parameter values calculated from analytical expressions (see Appendix~\ref{Ap:app3}) except for the nonlinearities of the confinement ($\kappa'_{ms}$, $\kappa'_{oe}$) and the damping $\xi$. This discrepancy may arise from the transition to a different dynamic regime in region~\textcircled{\raisebox{-1.0pt}{\hspace*{0.0pt}2}} that is not described by the model, or from a limitation in the theoretical predictions that takes into account nonlinearities only up to the second order.

With regard to the analytical model, we assert the advantages of vortex based STNO. The first is that the generation linewidth in the linear regime $\Delta\!f_{0}$ is of the order of 10~kHz instead of 1~MHz for uniform based STNO. Thus, vortex based STNOs present lower level of intrinsic phase noise. In addition, although the level of amplitude noise and the normalized dimensionless nonlinear frequency shift are of the same order of magnitude,  the conversion factor of amplitude-to-phase noise $\nu f_p=N p_0 /\pi$ (see Eq.~(\ref{eq:equ7})) is of the order of $10^4$ instead of $10^6$ for uniform based STNOs. It follows that the measured linewidth $\Delta\!f$ for vortex based STNOs is at least one order of magnitude smaller than uniform based STNO linewidth.

Moreover, we propose the possible improvements vortex based STNO features. Considering that the vortex based STNO linewidth $\Delta\!f$ is at least one order of magnitude higher than $2\Delta\!f_0$ at room temperature, the linewidth can be significantly reduced: By suppressing the amplitude-to-phase noise conversion, the phase noise would be equal to $2\Delta\!f_0$. In order to reduce the linewidth $\Delta\!f$, one can reduce the intrinsic linewidth $\Delta\!f_0$ or the amplitude-to-phase noise conversion (proportional to the amplitude noise level $S^{max}_{\delta\alpha}$ and the conversion factor of amplitude-to-phase noise $\nu f_p$) through the fabrication techniques increasing the spin polarization efficiency $a_j$ or decreasing the damping $D$ (see Eq.~(\ref{eq:equ40}), (\ref{eq:equ41}) and (\ref{eq:equ42})). To change $\nu$, there is a need of a modification of the magnetic disk shape and material as this term is link to the nonlinearities specific to the system. Furthermore, it is possible to decrease the phase noise with an external feed-back or excitation: as recently shown in experiments\citep{p._bortolotti_parametric_2013}, the parametrically excited gyrotropic motion demonstrates linewidths of the order of the intrinsic linewidth $\Delta\!f_0$. The origin of this low value may be the suppression of the amplitude-to-phase noise conversion.

\section*{Conclusions}
We investigate how noise affects the spin transfer induced vortex magnetization dynamics in a MTJ. From single shot time domain measurements, we measure the amplitude and phase noise associated to the gyrotropic motion of the vortex core for different values of external magnetic field and dc current. We propose a development of the nonlinear auto-oscillator model for the vortex based STNO where we define the amplitude and phase noise analytical expressions in function of the response-to-noise parameters $f_p$, $\Delta\!f_0$ and $\nu$. We show that the amplitude and phase noise are well described within the nonlinear auto-oscillator framework. From the comparison of the experiments to the analytical expressions we extract the response-to-noise parameters. Their values and evolutions with current are correlated to theoretical ones with good agreements apart for the nonlinearities. We deduce that the low linewidth of vortex based STNOs arises from the small value of amplitude-to-phase noise conversion factor $\nu f_p=N p_0 /\pi$ and generation linewidth in the linear regime $\Delta\!f_0$. The major noise contribution to the linewidth broadening is due to the amplitude-phase coupling and has to be suppressed if one wants to reduce the linewidth. 
 
\section*{acknowledgments}
The authors acknowledge A.S. Jenkins, R. Lebrun and A. Slavin for fruitful discussions, Y. Nagamine, H. Maehara and K. Tsunekawa of CANON ANELVA for preparing the MTJ films and the financial support from ANR agency (SPINNOVA ANR-11-NANO-0016) and EU FP7 grant (MOSAIC No. ICT-FP7- n.317950). E.G. acknowledges CNES and DGA for financial support.

\bibliography{MyLibrary}% Produces the bibliography via BibTeX.

\appendix{}
\section{Relation between the measured voltage, the vortex core position and their noise}
\label{Ap:app1}
The STNO is an oscillating resistance $R_s(t)$ where:
\begin{equation*}
R_s(t)=R_0+\Delta R(t)
\end{equation*}
with $R_0$ the resistance mean value and $\Delta R(t)$ the oscillating part.
Thus, the STNO is equivalent to the resistance $R_0$ in series with a voltage generator $e(t)=\Delta R(t) I$ (with $I$ the dc current injected in the STNO). 
The generator voltage that depends on the vortex core position can be expressed as:
\begin{equation*}
e(t)\,=\,\lambda s_0 \left(1+\frac{\delta s(t)}{s_0}\right) \cos (\dot{\theta}t+\delta\theta(t))
\end{equation*}
With a magnetoresistive factor $\lambda$ in which the influence of the perpendicular magnetic field $H_{\bot}$ on the magnetization of the layers and of the electrical current $I$ are also taken into account as:
\begin{equation*}
\lambda=\frac{I \Delta R_{p-ap}(I)}{2}\beta\sqrt{(1-(\frac{H_{\bot}}{4\pi M_S^{SAF}})^2)(1-(\frac{H_{\bot}}{4\pi M_S^{free}})^2)}
\end{equation*}
With the difference of resistance between the antiparallel state and the parallel state $\Delta R_{p-ap}(I)$ at a given $I$, the free layer saturation magnetization $M_S^{free}$, the SAF layer saturation magnetization $M_S^{SAF}$ and the conversion factor of the vortex core displacement into magnetization change\citep{guslienko_eigenfrequencies_2002} $\beta=\frac{2}{3}$ valid for displacements up to $60~\%$ of the dot radius.

The input measurement voltage $V(t)$ is expressed as:
\begin{equation*}
V(t)\,=\,V_0 \left(1+\delta\alpha(t)\right) \cos (2\pi ft+\delta\phi(t))
\end{equation*}
and the relationship between the measured voltage and the generator bias voltage is:
\begin{align*}
V(t)\,=\,\frac{R_{load}}{R_{load}+R_0}e(t) \\
\end{align*}
due to the impedance mismatch, where $R_{load}$ is the resistance input of the measurement device (for our experiments $R_{load}= 50\Omega$).

Finally we obtain:
\begin{align*}
V_0\,=\,\frac{R_{load}}{R_{load}+R_0}\lambda s_0 \\
\delta\alpha(t)\,=\,\frac{\delta s(t)}{s_0}\\
2\pi f\,=\,\dot{\theta}\\
\delta\phi(t)\,=\,\delta\theta(t)
\end{align*}
Thus the voltage normalized amplitude noise is equal to the magnetization normalized amplitude noise and the voltage phase noise is equal to the magnetization phase noise.

We express the measured integrated power $P_{int}$ as:
\begin{equation}
P_{int}\,=\,\frac{R_{load}}{(R_{load}+R_0)^2}{\langle V^2 \rangle}\,=\,\frac{R_{load}}{2(R_{load}+R_0)^2}(\lambda s_0)^2
\end{equation}

\section{Analytical expressions of the gyrotropic motion parameters}
\label{Ap:app3}
To express the influence of the perpendicular magnetic field $H_{\bot}$ on the free layer magnetization, we define:
\begin{equation*}
\theta_0\,=\,\arccos\left(\frac{H_{\bot}}{4\pi M_S^{free}}\right)
\end{equation*}
with $M_S^{free}$ the saturation magnetization of the free layer.
As already define in the reference~\citep{dussaux_field_2012} the spin transfer induced gyrotropic motion parameters are:
\begin{itemize}
\item[\textbullet]The vortex core magnetostatic confinement stiffness and its nonlinear coefficient:
\begin{align*}
\kappa_{ms}\,=\,\frac{10}{9}\mu_0 {M_S^{free}}^2\frac{L^2}{R}\sin^2\theta_0 \\
\kappa'_{ms}\,=\,0.25 \kappa_{ms}\sin^2\theta_0 
\end{align*}
\item[\textbullet]The vortex core confinement stiffness due to Zeeman interaction with the Oersted field and its nonlinear coefficient:
\begin{align*}
\kappa_{oe}\,=\,0.85C\mu_0 M_S^{free}LR\sin\theta_0\\
\kappa'_{oe}\,=\,-0.42C\mu_0 M_S^{free}LR\sin\theta_0
\end{align*}
\item[\textbullet]The gyrovector norm:
\begin{align*}
G\,=\,2\pi \frac{L M_S^{free}}{\gamma}(1-\cos\theta_0)\\
\end{align*}
\item[\textbullet]The damping coefficient:
\begin{align*}
D\,=\,2\pi\frac{\alpha\eta L M_S^{free}}{\gamma}\\
\eta\,=\,\left(\ln(R/4l_e)-\frac{1}{4}\right) \sin^2\theta_0
\end{align*}
\item[\textbullet]The spin transfer torque efficiency:
\begin{align*}
a_j\,=\,\pi \frac{\hbar P_{spin}}{2 |e|} p_z \sin^2\theta_0\\
p_z\,=\,\frac{H_{\bot}}{4\pi M_S^{SAF}}
\end{align*}
\end{itemize}
With $\mu_0$ the vacuum permeability, $L$ the thickness of the free layer, $R$ the radius of the free layer, $\gamma$ the gyromagnetic ratio, $\alpha$ the Gilbert damping, $l_e$ the exchange length, $\hbar$ the reduced Planck constant, $P_{spin}$ the spin polarization of the SAF layer, $e$ the electric charge and $M_S^{SAF}$ the SAF magnetization.

\begin{table}[h]
\caption{\label{tab:tabPar} Magnetic parameters used for theoretical estimations}
\begin{ruledtabular}
\centering
\begin{tabular}{cc} \toprule
		{$\alpha$} & $0.01$ \\
    {$M_S^{SAF}$} & $10.1\times10^{5}$ A.m$^{-1}$ \\
		{$M_S^{free}$} & $6.5\times10^{5}$ A.m$^{-1}$\\
		{$P_{spin}$} & $0.26$\\
		{$b$} & $33$ nm\\
\end{tabular}
\end{ruledtabular}
\end{table}

\section{Method to fit the experimental response-to-noise parameters}
\label{Ap:app2}
The response-to-noise parameters evolution are expressed in function of current $I$ as:
\begin{numcases}{} 
	\pi f_p \,= \,-\frac{D\kappa_{ms}}{G^2}+\left(\frac{a_j}{G\pi R^2}-\frac{D\kappa_{oe}}{G^2\pi R^2}\right)I \label{eq:equ50} \\
	2\Delta\!f_0 \, = \, \frac{k_BT}{\pi R^2}\frac{D}{G^2}\left(\frac{1}{p_0}+\frac{1}{\nu}\frac{G\kappa'_{ms}}{D\kappa_{ms}}h(I)\right) \label{eq:equ51} \\
	\nu \, = \, \frac{\zeta}{\frac{D}{G} (\zeta+\xi)}\, = \,\frac{G}{D} \frac{1}{1+\xi\frac{\kappa_{ms}}{\kappa'_{ms}}h(I)} \label{eq:equ52}
\end{numcases}
where:
\begin{equation}
h(I)\,=\,\frac{1+\kappa_{oe}I/\pi R^2\kappa_{ms}}{1+\kappa'_{oe}I/\pi R^2\kappa'_{ms}}
\end{equation}
and the carrier frequency as:
\begin{equation}
f_c\,=\,\frac{\kappa_{ms}}{2\pi G}\left((1+\frac{\kappa_{oe}I}{\kappa_{ms}\pi R^2})+\frac{\kappa'_{ms}}{\kappa_{ms}}p_0(1+\frac{\kappa'_{oe}I}{\kappa'_{ms}\pi R^2})\right)\label{eq:equ60}
\end{equation}

The experimental data are fitted with the previous equations to extract the gyrotropic motion quantities as the following:\\
\begin{itemize}
\item[\textbullet] We fit $f_p$ evolution with the Eq.(\ref{eq:equ50}) from which we extract $\frac{D\kappa_{ms}}{G^2}$ and $\frac{a_j}{G\pi R^2}-\frac{D\kappa_{oe}}{G^2\pi R^2}$.
\item[\textbullet] We fit the linear part of the frequency $f_c-\frac{\nu f_p}{2}$ from which we extract $\frac{\kappa_{ms}}{2\pi G}$ and $\frac{\kappa_{oe}}{\pi R^2 \kappa_{ms}}$. 
\item[\textbullet] We fit the frequency carrier with the Eq.(\ref{eq:equ60}) assuming the values of the parameters extracted above and injecting the oscillation power $p_0$. We deduce $\frac{\kappa'_{ms}}{\kappa_{ms}}$ and $\frac{\kappa'_{oe}}{\pi R^2 \kappa'_{ms}}$.
\item[\textbullet] We fit the evolution of $\nu$ with Eq.(\ref{eq:equ52}) when injecting the parameters above, from which we deduce $\frac{G}{D}$
\item[\textbullet] We fit the evolution of $\Delta\!f_0$ with Eq.~\ref{eq:equ51} knowing the parameters extracted above and injecting $\nu$ and the oscillation power $p_0$. From this we extract $\frac{k_BT}{\pi R^2}\frac{D}{G^2}$.
\end{itemize}

The oscillation power $p_0\,=\,{s_0}^2$ is evaluated with $s_0$ calculated as presented in Appendix \ref{Ap:app1}.

\end{document}